\documentclass[final,5p,times,twocolumn,number]{elsarticle}

\usepackage[T1]{fontenc}
\usepackage[utf8]{inputenc}

\makeatletter
\def\ps@pprintTitle{%
  \let\@oddhead\@empty \let\@evenhead\@empty
  \let\@oddfoot\@empty \let\@evenfoot\@oddfoot
}
\makeatother

\usepackage{booktabs}
\usepackage{array}
\newcolumntype{L}[1]{>{\raggedright\arraybackslash\hyphenpenalty=10000\exhyphenpenalty=10000}p{#1}}
\usepackage{amssymb}
\usepackage{amsmath}
\usepackage{graphicx}
\graphicspath{{figures/}{../figures/}{./}}
\usepackage{flushend}
\usepackage{stfloats}
\usepackage{float}
\usepackage{hyperref}

\hypersetup{colorlinks=true,linkcolor=blue,urlcolor=blue,citecolor=blue}
\newcommand{\code}[1]{\texttt{#1}}
\newcommand{\secref}[1]{\hyperref[#1]{Section~\ref*{#1}}}
\newcommand{\figref}[1]{Figure~\ref*{#1}}
\newcommand{\tabref}[1]{Table~\ref*{#1}}

\makeatletter
\renewcommand\subsection{\@startsection{subsection}{2}{\z@}%
           {12\p@ \@plus 6\p@ \@minus 3\p@}%
           {3\p@ \@plus 6\p@ \@minus 3\p@}%
           {\normalfont\normalsize\bfseries}}
\makeatother

\journal{arXiv}

\begin{document}

\begin{frontmatter}

\title{PaxosLease in Relativistic Inertial Frames}

\author{M\'arton Trencs\'eni (\code{mtrencseni@gmail.com})}

\begin{abstract}
A lease grants one holder exclusive authority over a resource for a fixed term, and its safety property guarantees that at most one holder has the lease at any time.  Under special relativity ``at any time'' has no frame-independent meaning for participants in relative motion.  This note first restates the classical safety property with light cones, requiring a holder's expiry to lie in the causal past of any later acquisition by a different holder, an order every observer agrees about.  Second, the paper modifies classical PaxosLease: the core change is that the containment and quarantine rules each gain a factor of $k = \sqrt{(1+\beta)/(1-\beta)}$, the relativistic Doppler factor, where $\beta$ is relative speed as a fraction of the speed of light.
\end{abstract}

\end{frontmatter}

\section{Introduction}\label{sec:intro}

\begin{figure*}[t]
\centering
\includegraphics[width=\textwidth]{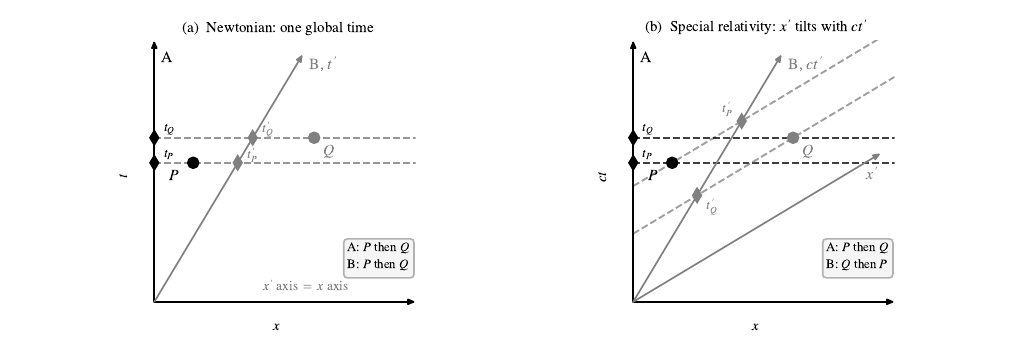}
\caption{\label{fig:simul}A is at rest and B moves at $\beta = 0.6$.  Dashed lines are lines of simultaneity, and the diamonds are the time coordinates of events $P$ and $Q$.  (a) Newtonian mechanics: observers A and B agree that $P$ occurs before $Q$, or $t_P < t_Q$ and $t'_P < t'_Q$.  (b) Special relativity: A still finds $t_P < t_Q$, but B's time coordinates are reversed, $t'_Q < t'_P$.}
\end{figure*}

A lease grants exclusive time-limited authority over a resource to one holder \cite{ref5}.  While the holder owns the lease, it acts on the resource without checking with the grantor(s).  Once the lease expires, the grantors may pass the lease to a new holder without hearing from the old one.  A holder that has crashed and stopped sending messages therefore loses the lease on schedule.  Lease algorithms have two important properties: (i) safety, that at most one holder has the lease at any time, and (ii) liveness, that a crashed holder does not block the distributed system.  All participants, and the resource itself, are modeled as point-like particles moving through spacetime.

The simplest approach to distributed algorithms is to assume an absolute frame of reference, with the single global time of Newtonian mechanics, where every event has a definite moment on the global timeline.  Distributed algorithms already handle failure modes commonly found in data centers on Earth: system clocks may be offset from one another, and may drift in rate by a bounded amount; messages may be delayed or lost; processes may crash and restart.  These faults limit what a participant can learn about events elsewhere, but none of them breaks the assumption of an absolute global time.

In special relativity there is no frame-independent shared \emph{now} for distant events.  An observer can nevertheless assign a time to a distant event relative to a chosen inertial frame: from the reception time of its light, subtract the light-travel time calculated in that frame.  Two observers at rest with respect to one another share an inertial frame and agree about when a distant event happened.  Two observers in relative motion generally do not.  The Lorentz transformation between their frames mixes space and time while preserving the same measured speed of light $c$ for both \cite{ref6}.  That invariant speed determines the geometry of a spacetime diagram.  With $x$ and $ct$ measured in the same units, light travels along the 45-degree diagonals for every inertial observer.  A moving observer's time axis and space axis both tilt by the same angle toward the light diagonal in the direction of motion.  The resulting change in what counts as a line of constant time is the geometric origin of the relativity of simultaneity.  For two spacelike separated events (meaning neither event can send a signal, traveling no faster than light, that reaches the other), two observers in relative motion may place them in opposite orders, as \figref{fig:simul} shows.  This is an aspect of our physical reality, described by special relativity; no better clock or measurement protocol removes this disagreement.

Let $P$ be one holder's lease expiring and $Q$ be the next, different holder acquiring it.  Safety demands that $P$ come first.  The liveness property requires that the acquisition not wait on a signal from $P$, since the first lease holder may crash right at $P$.  Classical lease algorithms' timing rules do not prevent these two events from being spacelike separated, and observers in certain inertial frames then date $P$ before $Q$ (the next lease starts after the previous expired), while others date $Q$ before $P$ (the next lease starts before the previous expired, violating the safety property).  Neither observer is mistaken; the ordering of spacelike separated events is relative to the observer.

In both Newtonian mechanics and special relativity, every observer agrees on cause and effect.  If a signal traveling no faster than light could have gone from one event to the other, the first could have caused the second, and every observer puts them in that order.  Two events at one object are the plainest case, since the object carries the record of the first forward to the second.  Relativity describes an object by its \emph{worldline}, the path it traces through space and time, and \emph{events along one worldline have one order for all observers}.\footnote{Lamport introduced \emph{happened-before} \cite{ref3} in 1978, with the analogy to relativity drawn explicitly.  The two are not identical, since relativity orders events that a signal could have connected while happened-before orders those that a message actually did.}

Motivating this study, consider a relay in deep space, built long ago and still running.  It executes any command sent to it.  It cannot tell who sent a command, cannot reject a command that a later one has superseded, and cannot arbitrate between two senders.  Several probes pass it on different trajectories at relativistic speeds, and each wants to use the relay to send messages into deep space.  This requires multiple commands, first to point the relay to the destination, and then to send the messages.  Two probes commanding it at once would leave it in a state neither intended.  The relay cannot stop two probes from commanding it at once, so the probes exclude one another instead: they run PaxosLease, and the lease holder alone commands the relay until its lease expires.

This note answers two questions.  How should the safety property be stated so that participants in relative motion can agree on whether it holds?  And what must PaxosLease \cite{ref1,ref2}, the lease protocol this note takes as its example, change so that it satisfies the restated property?

\secref{sec:paxoslease} recaps PaxosLease.  \secref{sec:exclusion} defines relativistic versions of safety, \secref{sec:rules} relativistic PaxosLease.  \secref{sec:discussion} discusses the results and their limitations.  Relativistic ideas are explained where they are first used.

\section{PaxosLease}\label{sec:paxoslease}

This section gives a brief description of PaxosLease, the lease protocol which we will adapt to special relativity in the later sections.  The protocol reuses the machinery of Paxos, ballots and prepare/accept rounds decided by majorities.  In Paxos, participants agree on an arbitrary value, while in PaxosLease agreement is reached on who holds the lease for some limited time, with automatic expiry; unlike Paxos, PaxosLease acceptors store nothing on disk.  The full algorithm is in \cite{ref1,ref2}.  The version modified here omits the optional early release of \cite{ref2}, which the argument of \secref{sec:containment} does not cover.

\begin{figure}[!ht]
\centering
\includegraphics[width=\columnwidth]{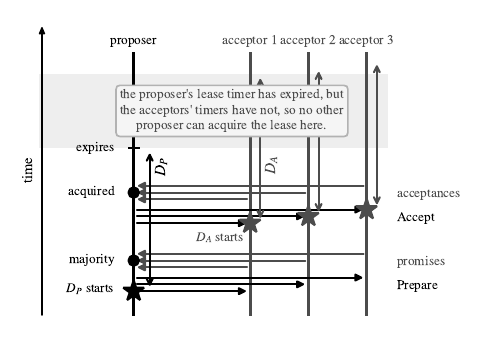}
\caption{\label{fig:flow}One acquisition in classical PaxosLease, as a message-sequence diagram rather than a spacetime diagram.}
\end{figure}

A proposer picks a fresh ballot, never used by any proposer, including its own earlier incarnations, starts a timer that expires $D_P$ later on its own clock, and sends \code{Prepare} to every acceptor.  An acceptor that answers promises to reject every smaller ballot and reports any live lease it has granted.  If a majority answer and none reports a live lease, or the only lease reported is the exact one this proposer still holds and is renewing, the proposer sends \code{Accept}.  An acceptor that accepts records the lease and reports it to any proposer that asks, until a deadline $D_A$ later on its own clock.  It will still accept a higher ballot from the same owner, which is a renewal, or from anyone once that deadline has passed, but it refuses an \code{Accept} that would replace a live lease of a different owner, for the reason given in \secref{sec:containment}.  An owner is a proposer incarnation, as in \cite{ref2}, so a restarted proposer is a new owner.  Once a majority of acceptances reach the proposer while its timer is still running, it has \emph{acquired} the lease and is its \emph{owner} until the timer, started at \code{Prepare}, expires.  Two proposers cannot both hold the lease because any two majorities share an acceptor, and that shared acceptor either reports the live lease or has already promised a ballot high enough to stop the lower-ballot attempt.  Acceptors keep promises and leases in memory only, so an acceptor that restarts has lost them, and it stays silent for a quarantine $D_Q$ before voting again.  \figref{fig:flow} shows one acquisition.

Throughout, $D_P$ is the proposer's lease duration, $D_A$ the acceptor's exclusion duration, $D_Q$ the restart quarantine, each measured on the clock of the participant that uses it, taken as ideal until the end of \secref{sec:quarantine}.\footnote{\emph{Proper time} is what an ideal clock carried along a worldline reads.  It is the only time a participant can measure directly.}   A relative speed is written $\beta$, as a fraction of the speed of light $c$.

\section{The safety property in relativistic inertial frames}\label{sec:exclusion}

\begin{figure*}[t]
\centering
\includegraphics[width=\textwidth]{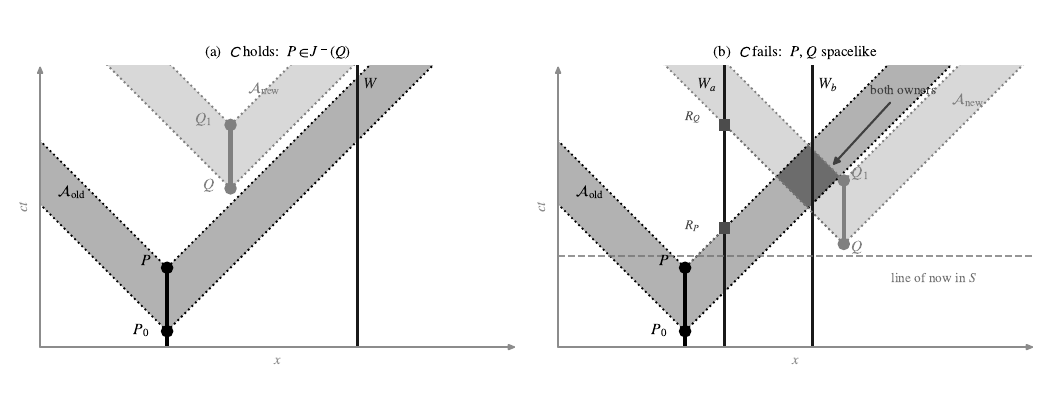}
\caption{\label{fig:cones}Authority regions $\mathcal{A}_{\mathrm{old}}$ and $\mathcal{A}_{\mathrm{new}}$, shaded, in the drawn frame $S$.  (a) Disjoint.  (b) Overlapping: $\mathsf{F}(S)$ holds, but $\mathsf{C}$ and $\mathsf{E}(W_b)$ fail.}
\end{figure*}

PaxosLease's safety property is that at most one proposer holds the lease at any time.  In special relativity ``at any time'' depends on the frame of reference, so the property is meaningless without choosing one, and different chosen frames may give different answers.  In \figref{fig:cones}(b), event $P$ is one owner's lease expiry and event $Q$ the next owner's acquisition; the two are spacelike separated, so observers in suitable frames disagree about their order.

\subsection{Exclusion definitions}

\textbf{Leases.}  Each successful attempt gives one lease, which runs along its owner's worldline from the acquisition to the expiry.  Participants are massive, so their worldlines are timelike, and an attempt is acquired strictly before its timer expires, so every lease has positive duration and its acquisition lies strictly inside the past cone of its expiry.  A renewal is a new lease by the same proposer, which usually overlaps the one it extends on the same worldline, so the three conditions below compare leases held by different proposers, and with more than two such leases each condition applies to every pair.

\textbf{Causal exclusion.}  Let $J^+(E)$ be the \emph{causal future} of an event $E$, the events a signal from $E$ traveling no faster than light can reach, and $J^-(E)$ its \emph{causal past}, the events from which such a signal can reach $E$.  These are the future and past light cones of $E$, their interiors, and $E$ itself.  Both sets are defined on spacetime itself rather than in any frame, so every observer agrees about them.  Causal exclusion $\mathsf{C}$ requires that of any two such leases, one has its expiry in the past cone of the other's acquisition; write $P$ for that expiry and $Q$ for that acquisition, so that
\[
  \mathsf{C}:\quad P \in J^-(Q) ,
\]
equivalently $J^+(Q) \subseteq J^+(P)$: everything the acquisition can reach, the expiry can reach too.  Every observer therefore agrees whether $\mathsf{C}$ holds.

\textbf{Resource exclusion.}  Each owner's lease runs between two events on its own worldline: the old owner acquires at $P_0$ and expires at $P$, and the new owner acquires at $Q$ and expires at $Q_1$.  An owner is authorized at the events where news of its acquisition could have arrived and news of its expiry could not, so the two regions are
\[
  \mathcal{A}_{\mathrm{old}} = J^+(P_0) \setminus J^+(P) , \qquad
  \mathcal{A}_{\mathrm{new}} = J^+(Q) \setminus J^+(Q_1) .
\]
Resource exclusion $\mathsf{E}(W)$ on a resource worldline $W$ requires the two regions to be disjoint along $W$,
\[
  \mathsf{E}(W):\quad W \cap \mathcal{A}_{\mathrm{old}} \cap \mathcal{A}_{\mathrm{new}} = \varnothing .
\]
This condition is frame independent as well, since it too is written with cones.

\textbf{Observer exclusion.}  Write $t_S(E)$ for the time coordinate of an event $E$ in a chosen inertial frame $S$.  Observer exclusion $\mathsf{F}(S)$ is classical PaxosLease's safety property as an observer at rest in $S$ reads it, and requires the two owners' authority intervals on that frame's time axis to be disjoint,
\[
  \mathsf{F}(S):\quad [\,t_S(P_0), t_S(P)\,) \cap [\,t_S(Q), t_S(Q_1)\,) = \varnothing .
\]

\begin{figure}[t]
\centering
\includegraphics[width=\columnwidth]{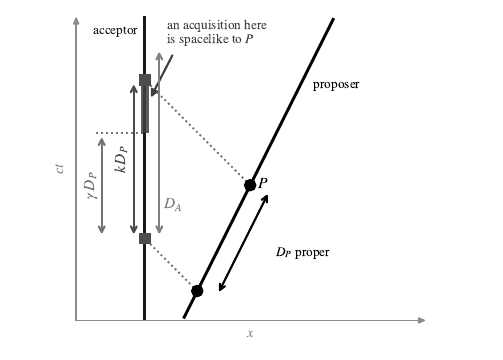}
\caption{\label{fig:kstretch}The proposer's timer runs $D_P$ of its own proper time, from \code{Prepare} to expiry.  The proposer recedes in between, so light from the two endpoints reaches the acceptor $kD_P$ apart.  The arrow marked $D_A$ must reach at least the second arrival.  An exclusion of $\gamma D_P$, the proposer's timer dilated into the acceptor's frame, would end $\beta\gamma D_P$ too early, and a proposer beside the acceptor that acquires in that gap does so at an event spacelike separated from the expiry $P$.}
\end{figure}

\subsection{Exclusion theorems}

Take any one resource worldline $W_0$ and any one frame $S$.  Since $\mathsf{C}$ names whichever owner is causally the earlier, in these labels it reads $Q \in J^+(P)$ or $P_0 \in J^+(Q_1)$.

\subsubsection{$\mathsf{C} \Leftrightarrow \mathsf{E}(W) \ \text{ for every } W$}
\textbf{Causal exclusion holds exactly when resource exclusion holds on every resource worldline.}  Under $\mathsf{C}$ the later owner's region lies inside the earlier owner's future cone, since $\mathcal{A}_{\mathrm{new}} \subseteq J^+(Q) \subseteq J^+(P)$, while $\mathcal{A}_{\mathrm{old}}$ excludes $J^+(P)$ by construction.  The two are disjoint everywhere, so no worldline meets both at one event, as \figref{fig:cones}(a) draws.  Conversely, let $\mathsf{C}$ fail, so neither $Q \in J^+(P)$ nor $P_0 \in J^+(Q_1)$, and compare the two acquisitions.  If $Q \in J^+(P_0)$, then $Q$ lies in both regions, being outside $J^+(P)$ by the first failure and outside $J^+(Q_1)$ because $Q_1$ is later along a timelike worldline; if $P_0 \in J^+(Q)$, then $P_0$ does the same by the mirror argument; and if the two are spacelike separated, their future cone surfaces meet at events null separated from each, which $J^+(P)$ and $J^+(Q_1)$, lying strictly inside those cones, cannot contain.  Every event lies on some timelike worldline, so a resource carried through the one found obeys both owners at once.

\subsubsection{$\mathsf{C} \Leftrightarrow \mathsf{F}(S) \ \text{ for every } S$}
\textbf{Causal exclusion holds exactly when observer exclusion holds in every frame.}  Causal order survives every boost, so $\mathsf{C}$ gives $t_S(P) \leq t_S(Q)$, or else $t_S(Q_1) \leq t_S(P_0)$, in every frame, and the interval that ends first ends before the other begins.  Conversely, each owner's acquisition precedes its expiry in every frame, and two such non-empty half-open intervals are disjoint exactly when one ends at or before the other begins, so $\mathsf{F}(S)$ says that $t_S(P) \leq t_S(Q)$ or $t_S(Q_1) \leq t_S(P_0)$.  No frame satisfies both, which would close the cycle $t_S(P_0) < t_S(P) \leq t_S(Q) < t_S(Q_1) \leq t_S(P_0)$, so the frames satisfying each form disjoint closed sets, and if $\mathsf{F}$ holds everywhere these sets cover the family of all frames, which is connected: one is empty and the other is everything.  A separation with non-negative time coordinate in every frame is future directed and causal, since a spacelike one reverses under some boost and a past directed one is negative everywhere.  The surviving case is therefore $Q \in J^+(P)$ or $P_0 \in J^+(Q_1)$, which is $\mathsf{C}$.

\subsubsection{$\mathsf{E}(W_0) \nRightarrow \mathsf{C}$}\label{sec:nimp1}
\subsubsection{$\mathsf{E}(W_0) \nRightarrow \mathsf{F}(S)$}
\textbf{Resource exclusion on one worldline does not imply causal exclusion or observer exclusion.}

\subsubsection{$\mathsf{F}(S) \nRightarrow \mathsf{C}$}
\subsubsection{$\mathsf{F}(S) \nRightarrow \mathsf{E}(W_0)$}\label{sec:nimp4}
\textbf{Observer exclusion does not imply causal exclusion or resource exclusion on a given worldline.}

\figref{fig:cones}(b) is the counterexample for \ref*{sec:nimp1}--\ref*{sec:nimp4}.  Causal exclusion fails there, yet the relay on $W_a$ satisfies resource exclusion, since light from $P$ reaches it at $R_P$ before light from $Q$ reaches it at $R_Q$, and the two leases are disjoint on the drawn frame's time axis, so $\mathsf{F}(S)$ holds.  A resource on $W_b$ obeys both owners at once, so $\mathsf{E}(W_b)$ fails, and since $\mathsf{C}$ fails, some frame violates observer exclusion by the equivalence above: for the drawn events, every observer moving from $P$'s side toward $Q$'s at between $0.15c$ and $0.95c$ sees the leases overlap.

\subsection{Commands and messages}

Assuming messages from the lease holder to the resource travel at the speed of light and are processed in arrival order, relativistic PaxosLease honoring causal exclusion yields correct execution at the resource.  The old owner sent its commands before $P$, so they arrive before light from $P$, and the new owner sends its commands at or after $Q$, so they arrive no earlier than light from $Q$.  Since $J^+(Q) \subseteq J^+(P)$, every command of the old owner arrives before any command of the new one.

\section{PaxosLease timeouts in relativistic inertial frames}\label{sec:rules}
In this section we will modify classical PaxosLease to achieve \emph{causal exclusion} in relativistic inertial frames by modifying timeouts, since requiring a release message before regranting would break the algorithm's liveness property.

\subsection{Relativistic Doppler factor}\label{sec:doppler}

Pick a frame and watch a clock move through it at a speed $\beta$.  The moving clock accumulates $1/\gamma$ seconds for every second of the frame's own time, where $\gamma = 1/\sqrt{1-\beta^2}$ is the Lorentz factor.  Since $0 \leq \beta < 1$, $\gamma \geq 1$ and $1/\gamma \leq 1$, so the moving clock runs slow.  That is time dilation.

An acceptor sees more than dilation, because it learns of each tick of the proposer's clock from the light that tick sends out, and if the proposer is receding, each successive tick is emitted from farther away.\footnote{The stretch depends on direction: $1/k$ for a proposer heading straight at the receiver, $\gamma$ for a tick emitted at closest approach, and $k$ for direct recession, the largest, so a single bound has to use $k$.}  In the acceptor's frame the ticks are $\gamma$ seconds apart, and in those $\gamma$ seconds the proposer moves $\beta\gamma$ light-seconds farther off, so the light of each tick has that much farther to travel than the light of the tick before.  The interval between arriving ticks is therefore stretched by
\[
  k = \gamma\,(1+\beta) = \sqrt{\frac{1+\beta}{1-\beta}}\,,
\]
the relativistic Doppler factor \cite{ref4}, and this is the factor $k$ the acceptor needs.  With $\beta \geq 0$ it satisfies $k = \gamma(1+\beta) \geq \gamma \geq 1$.  If the proposer's clock ticks once per second, the acceptor receives one tick every $k$ of its own seconds, so it observes the clock running at rate $1/k$.  The factor is computed from a relative speed, so it belongs to a pair of participants rather than to one.

An acceptor may use a pair's own $k$ if it knows a bound on their relative speed, for instance from trajectories both have committed to.  The Doppler shift it observes on the proposer's messages is no substitute: that shift depends on direction as well as speed, falls below one while the proposer approaches, and mixes the proposer's clock rate error into its apparent motion.  An acceptor without such a bound has to use $k_{\max}$, the stretch at an assumed bound $\beta_{\max}$ on relative speed, an assumption of the same kind as the clock drift bound that classical PaxosLease already makes.  So does any acceptor emerging from a restart, since it has forgotten which proposers it answered.  \secref{sec:containment} relies on the acceptor-side rule of \cite{ref2}, refusing to replace another owner's live lease.  With this approach the acceptors stay diskless.

\begin{table*}[tbp]
\centering
\small
\renewcommand{\arraystretch}{1.45}
\begin{tabular}{@{}L{0.23\textwidth}L{0.29\textwidth}L{0.42\textwidth}@{}}
\toprule
 & \textbf{Classical} & \textbf{Relativistic} \\
\midrule
Safety property & one holder per instant of a global time & $P \in J^-(Q)$ for leases of different proposers \\
Ballots, quorums, messages & as in classical PaxosLease & same, without early release \\
Proposer's lease timer & $D_P$ on its own clock & same \\
Acceptor's exclusion timer & $D_A \geq D_P\frac{1+\rho_{\max}}{1-\rho_{\max}}$ & $D_A \geq k\,D_P\frac{1+\rho_{\max}}{1-\rho_{\max}}$ \\
Acceptor's quarantine timer & $D_Q \geq D_P\frac{1+\rho_{\max}}{1-\rho_{\max}}$ & $D_Q \geq k_{\max}D_P\frac{1+\rho_{\max}}{1-\rho_{\max}}$ \\
Assumed bounds & maximum clock rate error $\rho_{\max}$ & $\rho_{\max}$, and maximum relative speed $\beta_{\max}$ \\
Acceptor uses & its own clock & same, and optionally a known bound on relative speed \\
Acceptor disk writes & none & none \\
\bottomrule
\end{tabular}
\caption{\label{tab:compare}Comparison of classical and relativistic PaxosLease.}
\end{table*}

\subsection{Containment}\label{sec:containment}
The containment rule requires the end of a granting acceptor's exclusion to lie in $J^+(P)$, the future cone of the expiry $P$ of the lease it granted.  Otherwise the acceptor could report the absence of a live lease before light from the expiry reached it, and the next proposer could acquire the lease outside $J^+(P)$.

Put the acceptor at rest and let the proposer recede, which is the worst case (\secref{sec:doppler}).  The proposer starts its attempt timer when it broadcasts \code{Prepare}, before it owns the lease.  If the acquisition succeeds, that same timer becomes the lease timer, and the lease expires when it runs out.  The broadcast and the expiry are $D_P$ apart on the proposer's clock, and light from each of them travels to the acceptor.  By the previous subsection the two arrivals are $kD_P$ apart, as \figref{fig:kstretch} shows.  An acceptor grants the lease only in response to messages of this attempt, sent by the proposer at or after its timer started, so the acceptor cannot start excluding before light from the broadcast could have reached it.  The acceptor's exclusion therefore starts no earlier than the first arrival, and if it lasts at least $kD_P$, it ends no earlier than the second arrival, which is when light from the expiry reaches the acceptor.  That is the containment rule with moving participants:
\[
  D_P \leq D_A / k .
\]
Nothing in the resulting rule depends on how far apart the participants are.

Causal exclusion follows from the containment rule, provided an acceptor also keeps the leases it has granted.

\textbf{Lemma (lease retention).}  Once an acceptor has accepted a lease, and while it stays up, it reports a live lease of that lease's owner at least until its exclusion for that lease ends, which by the containment rule is no earlier than when light from the lease's expiry reaches it.  Three rules guarantee this.  First, an acceptor never replaces a live lease of a different owner, even under a higher ballot, the acceptor-side rule of \cite{ref2}.  Second, an \code{Accept} from the same owner, a renewal, may replace the record, but its new deadline is the later one, being the same duration from a later moment.  Third, there is no early release: the optional release of \cite{ref2} clears the record and would break the property.

\textbf{Theorem (causal exclusion).}  For any two leases held by different proposers, one lease's expiry lies in the past cone of the other's acquisition.

\textbf{Proof.}  Write $P$ for the expiry of the lower-ballot lease and $Q$ for the acquisition of the higher-ballot one.  The majority whose answers to the higher \code{Prepare} were counted shares an acceptor with the lower-ballot lease's accept majority.  Consider that acceptor's acceptance of the lower ballot and its answer to the higher \code{Prepare}.

\textbf{Case 1: no restart between the two events.}  The acceptor accepted first: a prior promise of the higher ballot would have made it refuse the lower \code{Accept}.  Both events lie on its worldline, so every observer agrees on their order.  By the lemma, until its exclusion for the lower-ballot lease ended it would have reported a live lease of that owner, which the higher-ballot proposer cannot count as open, whether acquiring afresh or renewing, since the renewal qualifier admits only its own lease.  So the answer came after the exclusion ended, and hence after light from $P$ reached the acceptor.  The acquisition $Q$ follows the answer, so $P \in J^-(Q)$.

\textbf{Case 2: restart after the acceptance, before the answer.}  The acceptor forgot its acceptance of the lower-ballot lease.  The quarantine of \secref{sec:quarantine} places its answer, and hence $Q$, in the future cone of $P$.

\textbf{Case 3: restart after the answer, before the acceptance.}  The acceptor forgot its promise of the higher ballot and then accepted the lower one.  The same quarantine places that acceptance, and hence the lower-ballot acquisition, in the future cone of the higher-ballot expiry.  Here the lower-ballot lease is the causally later one, and causal exclusion still holds.

\subsection{Quarantine}\label{sec:quarantine}

An acceptor that restarts has forgotten every promise it made and every lease it granted, so it quarantines itself for $D_Q$ time, until light from the expiry of any previous proposer leases has reached it.  Each one belongs to an attempt one of whose messages, sent at or after its \code{Prepare}, reached the acceptor before the crash, so light from that \code{Prepare} had too, and light from the attempt's expiry arrives at most $k_{\max}D_P$ later.  The acceptor no longer knows which proposers those attempts belonged to, so no known pair value is available to it.  Waiting $k_{\max}D_P$ after restarting therefore places the acceptor in $J^+(P)$ for the expiry $P$ of every attempt it has forgotten:
\[
  D_Q \geq k_{\max}\,D_P .
\]
A \code{Prepare} that arrives during the quarantine needs no such argument, since the acceptor does not answer it.

With imperfect clocks, each clock's rate against its own proper time lies between $r_{\min} = 1-\rho_{\max}$ and $r_{\max} = 1+\rho_{\max}$, for $0 \leq \rho_{\max} < 1$; containment and quarantine then require, respectively,
\[
  k\,D_P\,r_{\max} \leq D_A\,r_{\min},
  \qquad
  k_{\max}\,D_P\,r_{\max} \leq D_Q\,r_{\min}.
\]

\section{Discussion}\label{sec:discussion}\label{sec:limitations}

The paper's results have two limitations.  First, every command must travel at the speed of light, and the resource must act on commands in the order they arrive; those arrivals lie on the resource's own worldline, so their order is one every observer agrees about.  Second, the note uses special relativity only, which assumes flat spacetime, a good approximation far from large masses and a poor one near them, and it assumes that all participants move inertially.  \tabref{tab:compare} summarizes how the rules change from classical to relativistic PaxosLease.


\end{document}